\documentclass[
 reprint,
 amsmath, amssymb,
 aps,
 prl,
 superscriptaddress
]{revtex4-1}

\usepackage{graphicx}
\usepackage{color} 

\begin{document}


\title{Connecting the Kuramoto model and the chimera state}

\author{Tejas Kotwal}
\affiliation{
Department of Mathematics, 
Indian Institute of Technology Bombay,
Mumbai, India
}

\author{Xin Jiang}
\affiliation{
LMIB and School of Mathematics and Systems Science,
Beihang University,
Beijing, China,
}

\author{Daniel M. Abrams}
\affiliation{
Department of Engineering Sciences and Applied Mathematics;
Department of Physics and Astronomy,
Northwestern University,
Evanston, IL, USA,
}


\begin{abstract}
Since its discovery in 2002, the chimera state has frequently been described as a counterintuitive, puzzling phenomenon. The Kuramoto model, in contrast, has become a celebrated paradigm useful for understanding a range of phenomena related to phase transitions, synchronization and network effects. Here we show that the chimera state can be understood as emerging naturally through a symmetry-breaking bifurcation from the Kuramoto model's partially synchronized state. Our analysis sheds light on recent observations of chimera states in laser arrays, chemical oscillators, and mechanical pendula.
%
\end{abstract}

\maketitle

\pagebreak

The chimera state \cite{Kuramoto2002} was so dubbed \cite{Abrams2004} because of its similarity to the Greek mythological creature made up of parts (a lion's head, a goat's body, a serpent's tail) that didn't belong together.  It seemed unbelievable that identical oscillators, coupled in identical ways to their neighbors, could behave in radically different fashions.  Appeals to insight from other symmetry-breaking phenomena were fruitless, because in most comparable problems the symmetric state loses stability; here, both the symmetric (fully-synchronized) and the asymmetric (chimera) state were simultaneously stable.

Despite continued research on chimera states (see, e.g., \cite{Montbrio2004, Sakaguchi2006, Abrams2008, Arenas2008, Bordyugov2010, Motter2013, Omelchenko2013, Panaggio2013, GONZALEZAVELLA2014, Rothkegel2014, Ashwin2015, Bastidas2015, Panaggio2015_1, Panaggio2015_2, Nishikawa2016, Roulet2016, Ulonska2016, Cho2017}) and significant mathematical insight, they have resisted intuitive explanation. In this Letter, we show how intuition can indeed yield understanding of the chimera state.  For a natural extension of the model, it occurs as the limiting case of a pitchfork bifurcation that destabilizes the symmetric state.


\textbf{Mathematical background.}
The ``traditional'' Kuramoto model \cite{Kuramoto1975, Kuramoto1984book, Kuramoto1984} has been extensively studied (see, e.g., \cite{Acebron2005, STROGATZ2000}); it is
\begin{equation}  \label{eq:KM0}
  \dot{\theta}_i = \omega_i - \frac{K}{N} \sum_{j=1}^N \sin(\theta_i - \theta_j),
\end{equation}
where $\theta_i$ and $\omega_i$ are the phase and the natural frequency of the $i$th oscillator in a population of $N$ coupled oscillators. Typically the natural frequencies $\left\{\omega_i\right\}$ are drawn from a known distribution $g(\omega)$.

In the thermodynamic limit $N \to \infty$, the continuum of oscillators at each $\omega$ value can be described by the probability density function $f(\theta, t; \omega)$, which must satisfy the continuity equation. The sum in \eqref{eq:KM0} represents an average of the sine of the phase difference over all oscillators, and is therefore generalized as an integral.  Thus the instantaneous velocity of an oscillator with natural frequency $\omega$ becomes
\begin{align} \label{eq:KM1}
 v(\theta, t; &\omega) = \omega - \\ \nonumber
 & K \int_{-\pi}^{\pi} \int_{-\infty}^{\infty} \sin \big( \theta - \theta ' \big) \ f \big( \theta ', t; \omega ' \big) \ g \big( \omega ' \big) \ d \omega ' \ d \theta '.  
\end{align}

A simple system that can form chimera states consists of two clusters of oscillators \cite{Abrams2008, Laing2009, Laing2012_2}, with equations given by
\begin{equation} \label{eq:twoclusterode}
  \dot{\theta}_{i}^{\sigma} = \omega_{i}^{\sigma} - \sum_{\sigma' = 1}^{2} \frac{K_{\sigma \sigma'}}{N_{\sigma'}} \sum_{j=1}^{N_{\sigma'}} \sin(\theta_i^{\sigma} - \theta_j^{\sigma'} + \alpha).
\end{equation}
Here the two clusters are identified by $\sigma \in \{1,2\}$, $\omega_{i}^{\sigma}$ are drawn from a distribution $g ( \omega )$, $N_{\sigma}$ is the number of oscillators in cluster $\sigma$ and $\alpha$ is the phase lag.  The coupling strength between oscillators in cluster $\sigma'$ and those in cluster $\sigma$ is given by $K_{\sigma \sigma'}$; we take $K_{11} = K_{22} = \mu > 0$, and $K_{12} = K_{21} = \nu > 0$, with $\mu > \nu$ (so intra-cluster coupling is stronger than inter-cluster coupling). By rescaling time, we may set $\mu + \nu = 1$. It is useful to define the parameters $A = \mu - \nu$ and $\beta = \pi/2 - \alpha$, because, as will be shown later, chimera states exist in the limit where these quantities are small. 

We begin by analyzing system \eqref{eq:twoclusterode} in the continuum limit where $N_{\sigma} \rightarrow \infty$ for $\sigma \in \{1,2\}$. Two probability densities $f^{\sigma}(\theta, t; \omega)$ are assumed to exist and satisfy continuity equations for each population. Thus, equations \eqref{eq:twoclusterode} become
\begin{multline}
  v^{\sigma}(\theta, t; \omega) = \omega \ -  \sum_{\sigma ' = 1}^{2} K_{\sigma \sigma '} \int_{-\infty}^{\infty} \int_{-\pi}^{\pi} \sin\big( \theta - \theta ' + \alpha \big) \\
  \times f^{\sigma '}(\theta ', t; \omega ') \ d \theta ' \ d \omega ',
\end{multline}
where $v^\sigma$ represents the phase velocity $\dot{\theta}$ of oscillators in cluster $\sigma$.  Note that we have dropped the superscripts on $\theta$ and $\omega$ to ease the notation---$\theta$ means $\theta^{\sigma}$ and $\theta '$ means $\theta^{\sigma '}$, and similarly for $\omega$.  

We define a complex order parameter for each cluster
\begin{equation}
  z_{\sigma}(t) = \left<e^{i \theta^\sigma} \right> = \int_{-\infty}^{\infty} \int_{-\pi}^{\pi} e^{i \theta} \ f^{\sigma}(\theta, t; \omega) \ d \theta \ d \omega,
\end{equation}
so that $v^{\sigma}(\theta, t; \omega)$ simplifies to
\begin{equation}
  \small
  v^{\sigma}(\theta, t; \omega) =  \omega + \frac{1}{2i} \sum_{\sigma ' = 1}^{2} K_{\sigma \sigma '} (z_{\sigma '} e^{-i(\theta + \alpha)} - \overline{z}_{\sigma '} e^{i(\theta + \alpha)}).
\end{equation}

Ott and Antonsen proposed the following ansatz for the expansion of $f^{\sigma}(\theta, \omega, t)$ in a Fourier series \cite{Ott2008}:
\begin{equation} \label{eq:Ottansatz}
  f^{\sigma}(\theta, t; \omega) = \frac{g ( \omega )}{2\pi} \Bigg( 1 + \bigg( \sum_{n = 1}^{\infty} (a_{\sigma}(\omega, t) e^{i \theta})^{n} + \text{c.c.} \bigg) \Bigg),
\end{equation}
where $\text{c.c.}$ stands for complex conjugate. Plugging \eqref{eq:Ottansatz} into the continuity equation yields a system of two coupled partial integro-differential equations
\begin{equation} \label{eq:ampeqn}
  \frac{\partial a_{\sigma}}{\partial t} + i \omega a_{\sigma} + \frac{1}{2} \sum_{\sigma ' = 1}^{2} K_{\sigma \sigma '} (z_{\sigma '} a_{\sigma}^{2} e^{-i\alpha} - \overline{z}_{\sigma '} e^{i\alpha}) = 0,
\end{equation}
where $z_{\sigma} (t) = \int_{-\infty}^{\infty} g ( \omega ) \ \overline{a}_{\sigma}(\omega, t) \ d \omega$. 

We henceforth take $g ( \omega )$ to be a Lorentzian (Cauchy) distribution with mean zero and scale parameter (width) $D$, so $\pi g(\omega) = D / (\omega^{2} + D^{2})$. This allows $z_{\sigma}(t)$ to be evaluated analytically by contour integration, yielding $z_{\sigma} (t) = \overline{a}_{\sigma} (-iD, t)$; plugging this into equation \eqref{eq:ampeqn} results in a two-dimensional system of complex ordinary differential equations (ODEs) that describe the dynamics of the order parameters of the two clusters. 

We rewrite the ODEs in polar coordinates by substituting $z_{1} = r_{1} e^{-i \phi_{1}}$ and $z_{2} = r_{2} e^{-i \phi_{2}}$ and defining $\phi = \phi_{1} - \phi_{2}$. This yields the three-dimensional system of real ODEs
\begin{subequations} \label{eq:zdyn}
\small
	\begin{align} 
		\dot{\phi} &= \bigg( \frac{1 + r_{1}^{2}}{2r_{1}}\bigg) [\mu r_{1} \sin \alpha - \nu r_{2} \sin (\phi - \alpha)] \nonumber \\
		 &\quad - \bigg( \frac{1 + r_{2}^{2}}{2r_{2}}\bigg) [\mu r_{2} \sin \alpha + \nu r_{1} \sin (\phi + \alpha)]. \label{subeq:phidyn} \\
		\dot{r}_{1} &= -D r_{1} + \bigg( \frac{1 - r_{1}^{2}}{2}\bigg) [\mu r_{1} \cos \alpha + \nu r_{2} \cos (\phi - \alpha)] \label{subeq:r1dyn} \\
		\dot{r}_{2} &= -D r_{2} + \bigg( \frac{1 - r_{2}^{2}}{2}\bigg) [\mu r_{2} \cos \alpha + \nu r_{1} \cos (\phi + \alpha)]  \label{subeq:r2dyn}
	\end{align}
\end{subequations}

For our analysis, we set $\mu = (1 + A)/2$, $\nu = (1 - A)/2$ and $\alpha = \pi/2 - \beta$. The bifurcation analysis will be carried out in the three-dimensional parameter space $(\beta, A, D)$.

\textbf{Searching for the connection.}
In previous works \cite{Abrams2008, Montbrio2004, Abrams2004}, the $D = 0$ case was analyzed, and it was shown that the chimera state disappears via a saddle-node bifurcation with an unstable saddle chimera state. There was no apparent connection between the stable chimera state and the fully synchronous state---both states seemed to be stable with different basins of attraction \cite{Martens2016}. In this section, we attempt to find such a connection by searching the 3-D parameter space $(\beta, A, D)$.

\begin{figure}[thb]
    \centering
	\vspace{0mm}
    \includegraphics[width=\columnwidth]{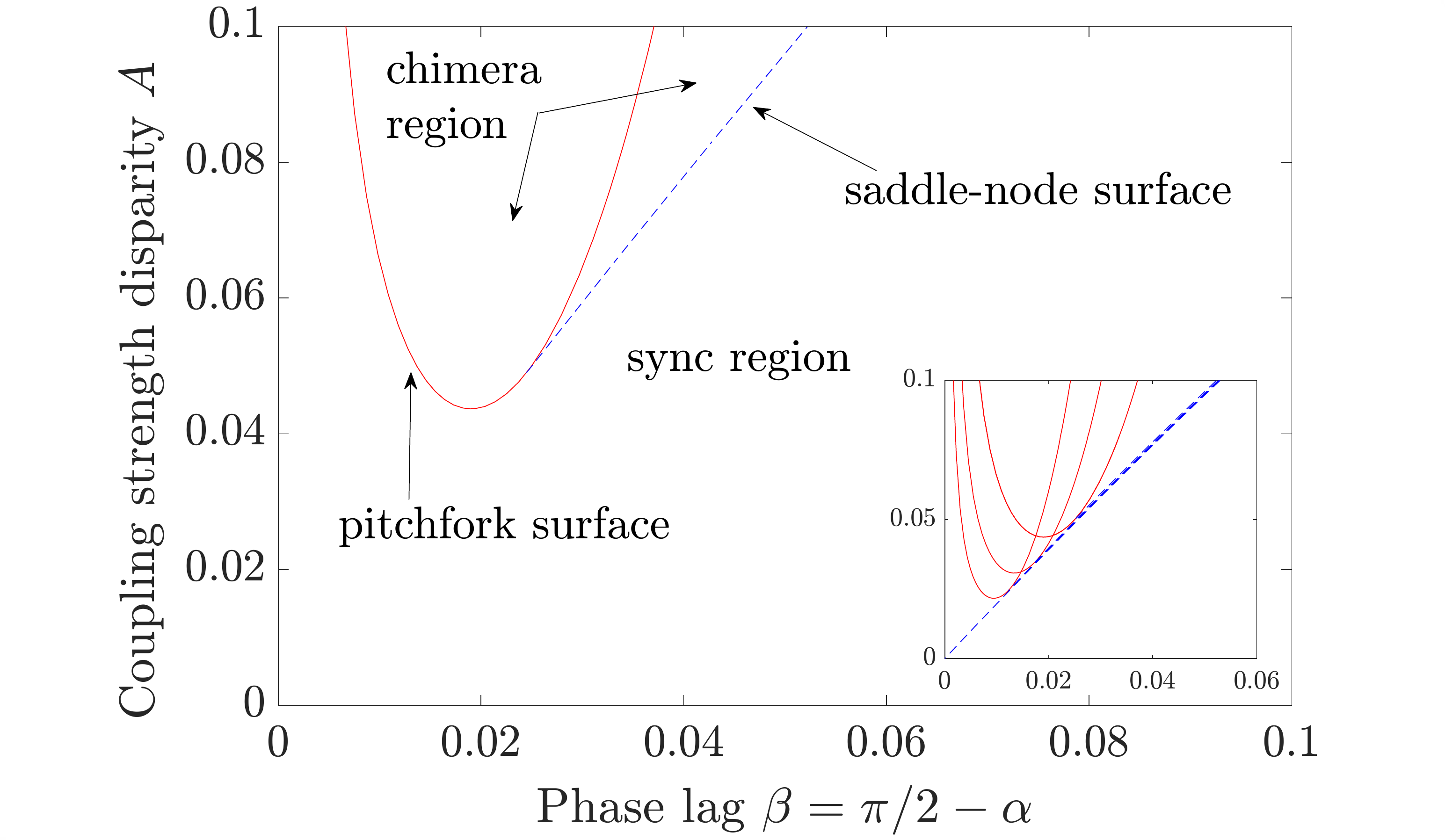}
    \vspace{-7mm}
    \caption{Bifurcations of equilibria from Eqs.~\eqref{eq:zdyn} with $D = 0.0006$, $\mu = (1+A)/2$, $\nu=(1-A)/2$.  Red solid curves: pitchfork bifurcation; blue dashed curves: saddle-node bifurcation.  Inset shows how curves shift as parameters change, with $D=0$, $D=0.00015$, $D=0.0003$, and $D=0.0006$ from bottommost curve to topmost (note that pitchfork bifurcation does not occur with $D=0$; in that case saddle-node curve extends to origin). Via numerical continuation \cite{Ermentrout2002, Doedel2007}. }
    \label{fig:Dslicechimerawedge}
\end{figure}

\setlength{\belowcaptionskip}{-10pt}
\begin{figure}[thb]
    \centering
	\vspace{-4mm}
    \includegraphics[width=\columnwidth]{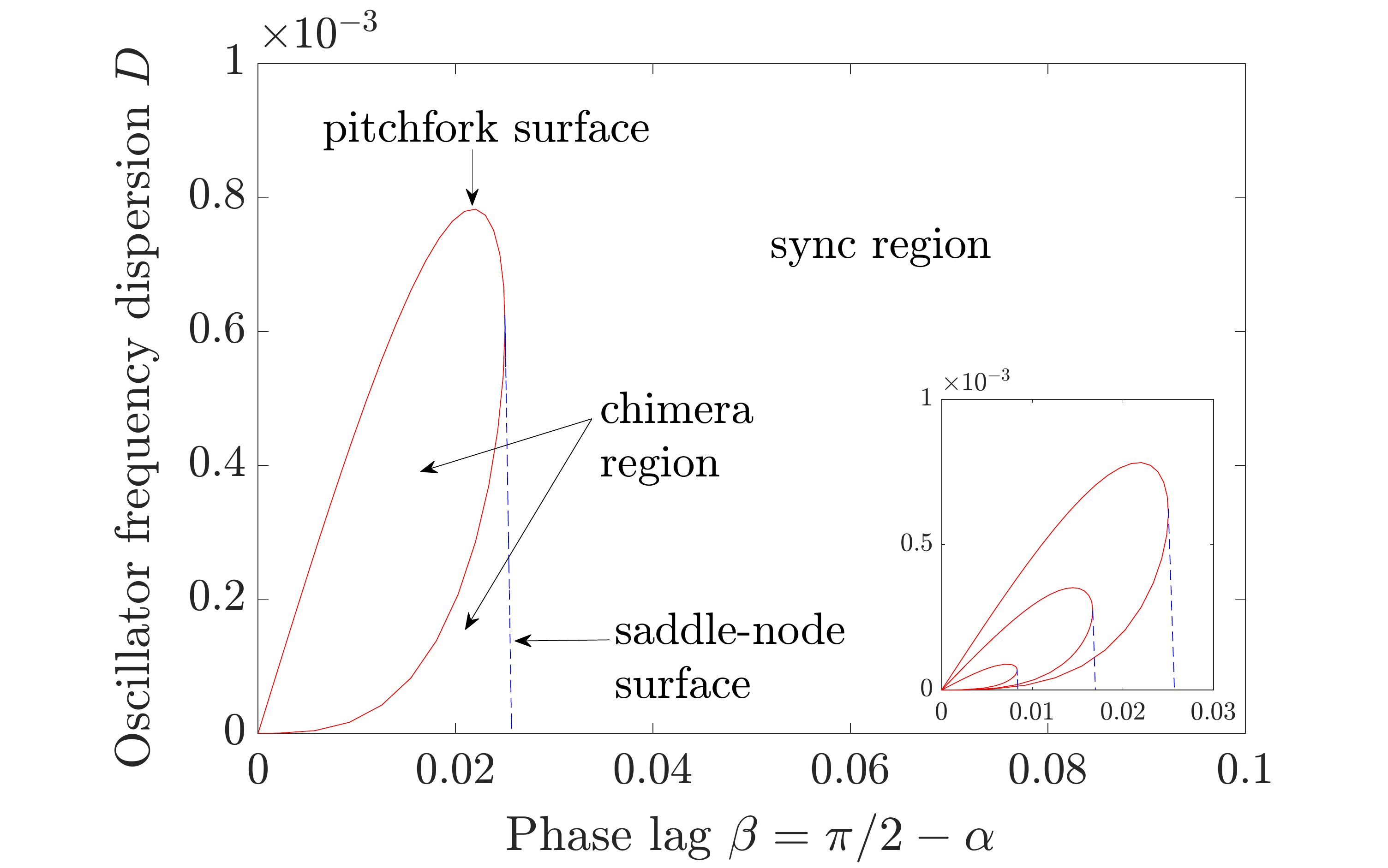}
    \vspace{-7mm}
    \caption{Bifurcations of equilibria from Eqs.~\eqref{eq:zdyn} with $A = 0.05$ ($\mu = 0.525$, $\nu=0.475$).  Red solid curves: pitchfork bifurcation; blue dashed curves: saddle-node bifurcation.  Note the peculiar balloon-like shape of this section originating from the $A$ axis.  Inset shows how curves shift as parameters change, with $A=0.0167$, $A=0.033$, and $A=0.05$ from bottommost curve to topmost. Via numerical continuation \cite{Ermentrout2002, Doedel2007}. }
    \label{fig:Aslicechimerawedge}
\end{figure}


In Fig.~\ref{fig:Dslicechimerawedge} we show bifurcation curves in $A$ vs $\beta$ for various slices of $D \ge 0$. For $D>0$, we observe a pitchfork bifurcation curve close to the origin that does not exist in the $D=0$ case. We also observe that the saddle-node curve does not extend to the origin for $D>0$. Another cross-sectional view with fixed $A$ is shown in Fig.~\ref{fig:Aslicechimerawedge}, where the pitchfork surface appears ``balloon-like.''  Laing obtained a similar figure in \cite{Laing2009}, where a chimera state on a 1-D ring with dispersion in natural frequency was analyzed.
\begin{figure}[thb]
    \centering
    \vspace{-7mm}
    \includegraphics[width=0.9\columnwidth]{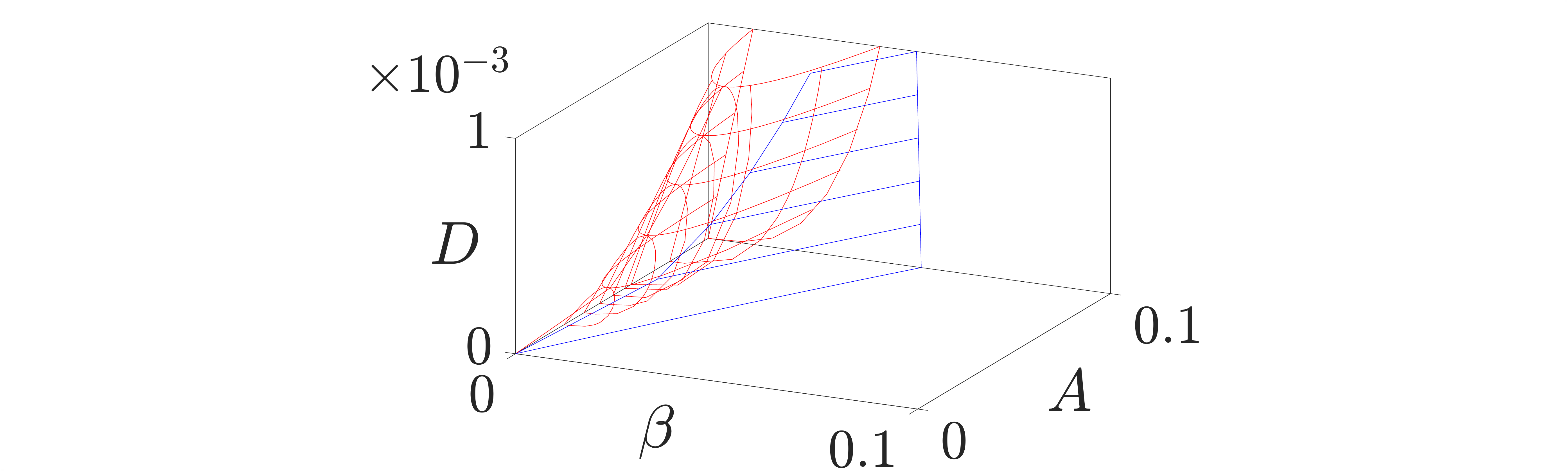}
    \vspace{-4mm}
    \caption{Three-dimensional bifurcation surfaces for equilibria of Eqs.~\eqref{eq:zdyn}. Red curved surface: pitchfork bifurcation; blue planar surface: saddle-node bifurcation. Via numerical continuation  \cite{Ermentrout2002,Doedel2007}.} 
    \label{fig:chimera_wedge}
\end{figure}
%
In Fig.~\ref{fig:chimera_wedge} we assemble a set of two-parameter sections to construct a 3-D bifurcation plot 
%
\footnote{All bifurcation diagrams were plotted using either XPP-Aut or our own continuation code. XPP-Aut \cite{Ermentrout2002} is software that consists of XPP, a differential equation solver, and AUTO \cite{Doedel2007}, an engine for numerical continuation of bifurcations. }.

\textbf{Understanding the chimera ``wedge.''}
How does the geometry of the bifurcation surfaces affect the order parameters $r_{1}, r_{2}$ of the two clusters? We address this question by looking at the effect of frequency dispersion on $\beta$-parameter sweeps of $r_{1}, r_{2}$. 


\begin{figure}[thb]
    \centering
	\vspace{0mm}
    \hspace{2.25mm}
   	\includegraphics[width=0.763\columnwidth]{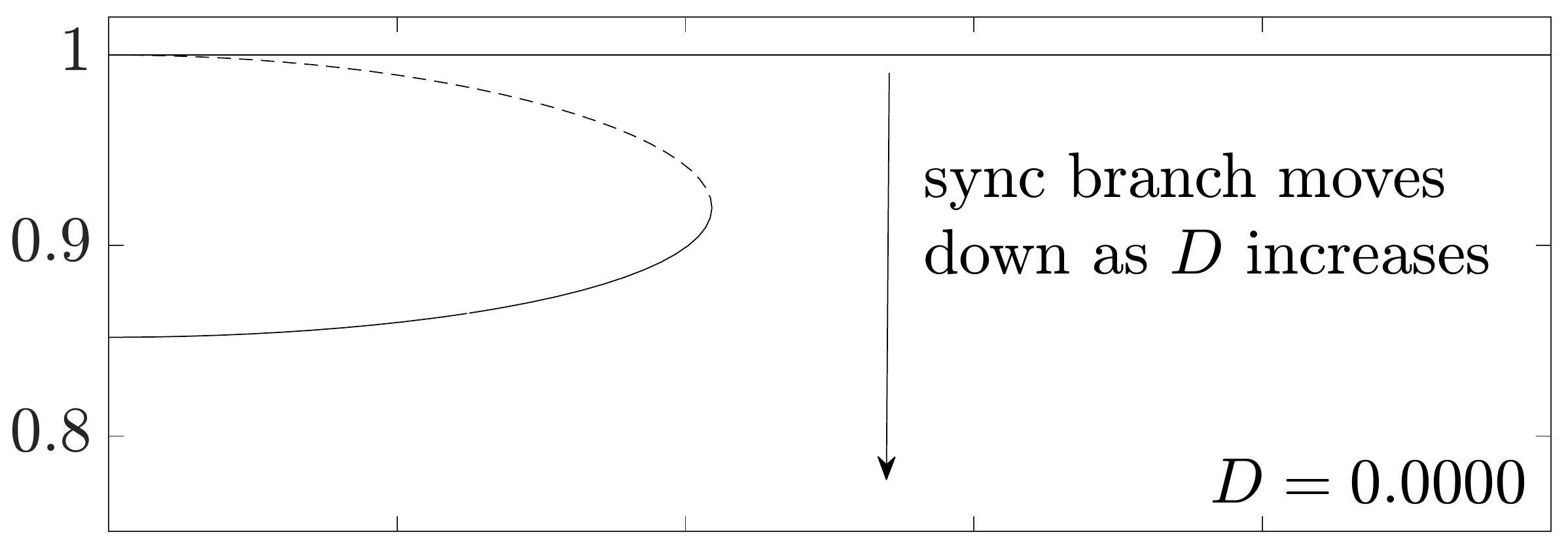}
   	~\\
	\hspace{2.25mm}
   	\includegraphics[width=0.763\columnwidth]{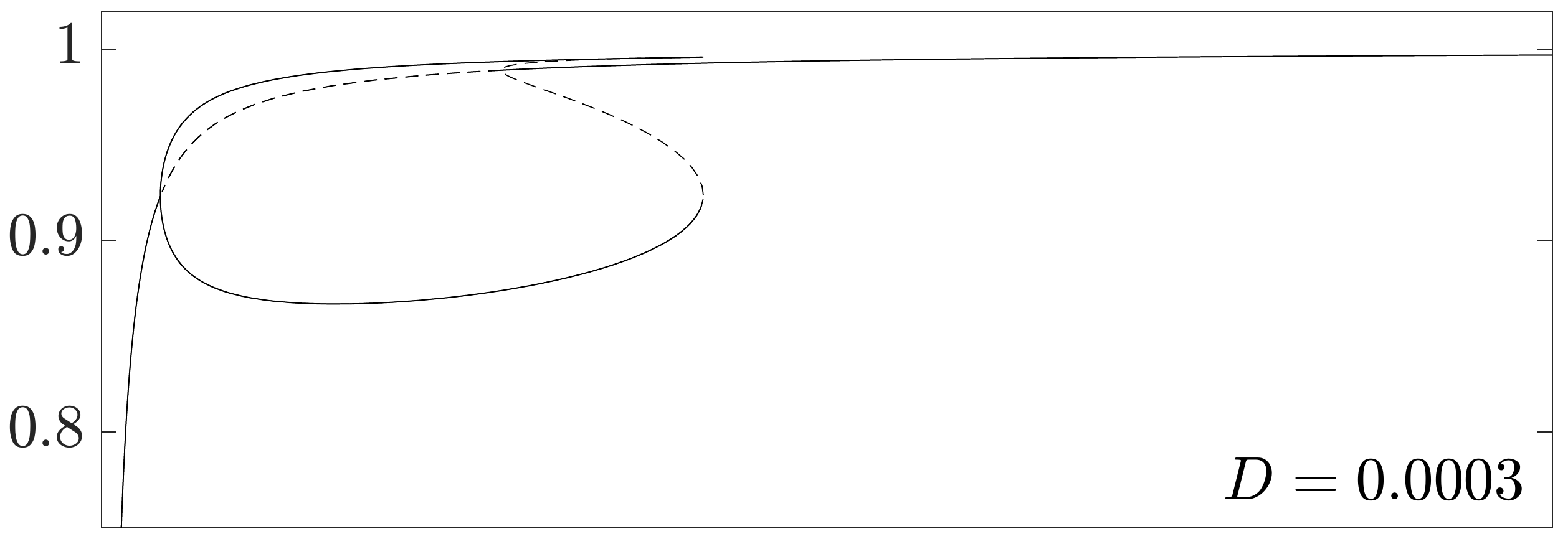}
   	~\\
	\hspace{-1mm}
   	\includegraphics[width=0.8\columnwidth]{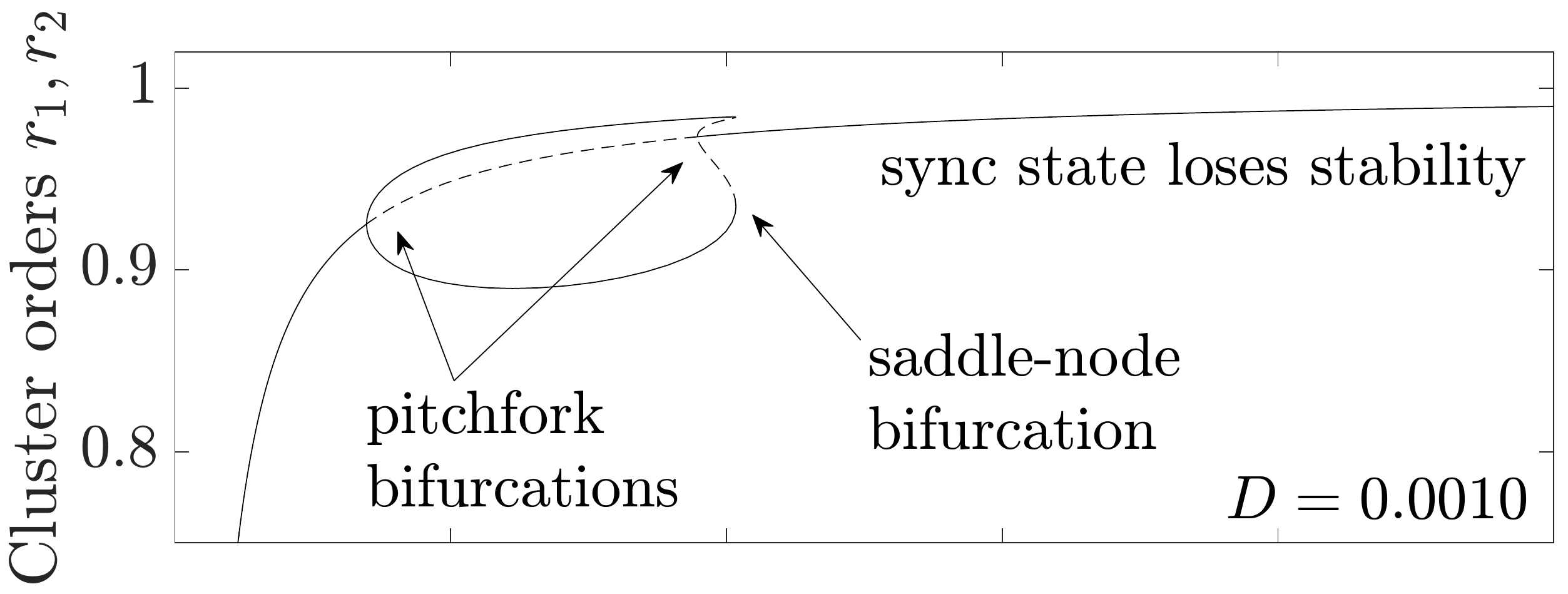}
   	~\\
	\hspace{3.5mm}
   	\includegraphics[width=0.781\columnwidth]{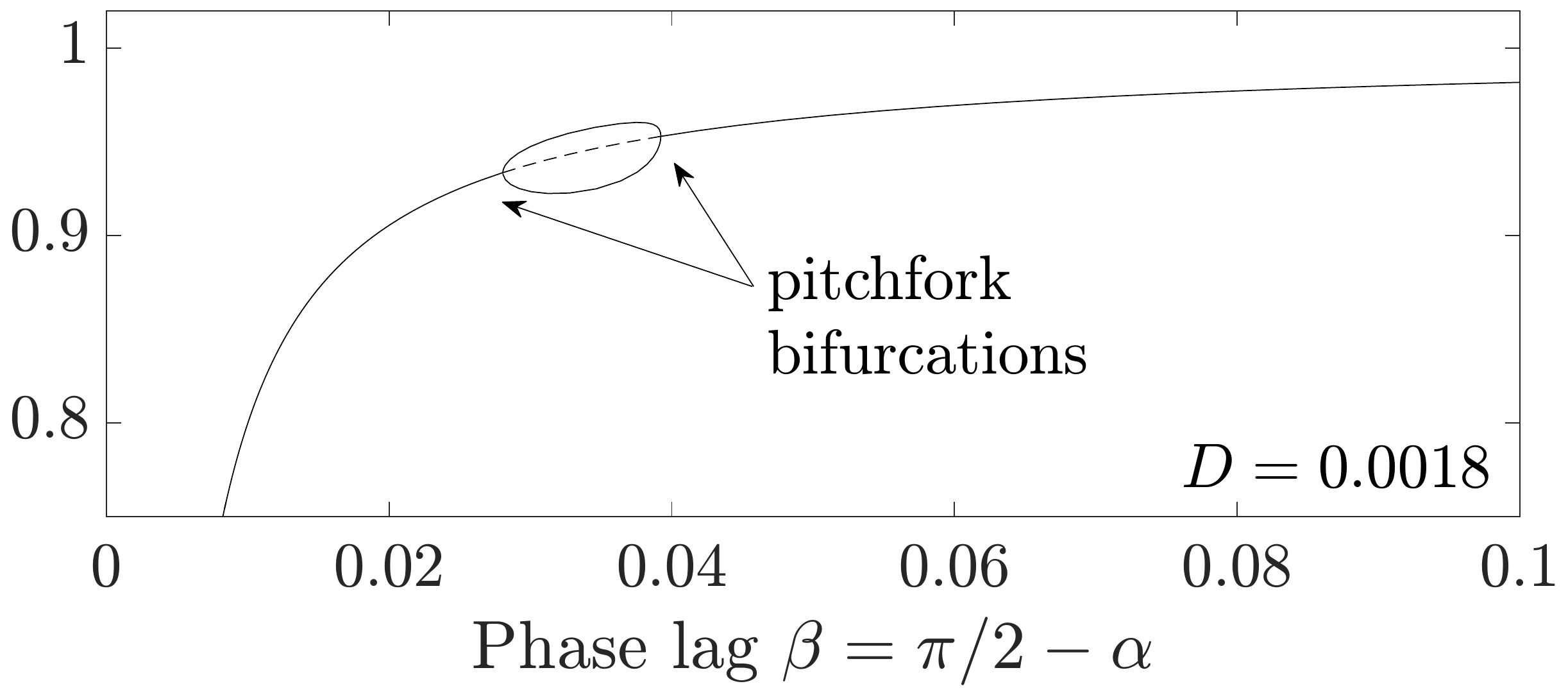}
	\vspace{-4mm}
    \caption{Cluster order parameter $r_1$ and $r_2$ vs.~phase lag $\beta$ for three values of $D$ and $A=0.08$ Solid curves: stable branches; dashed curves: unstable branches. Top panel: symmetric sync state has $r_1=r_2=1$ and chimera states have $\{r_1=1, r_2<1\}$ or $\{r_1<1, r_2=1\}$. Middle and bottom panels: branches above and below central branch correspond to pairs of chimera states symmetric under interchange of cluster number, i.e., $\{r_1=a, r_2=b\}$ and $\{r_1=b, r_2=a\}$.  Center branch always corresponds to symmetric extension of sync state with $r_1=r_2$. Via numerical continuation \cite{Ermentrout2002, Doedel2007}.
}
    \label{fig:bifurcationasDinc}
\end{figure}

For the $D = 0$ case, only a saddle-node bifurcation exists \cite{Abrams2008}. As $D$ is increased from $0$, we expect the order parameter of the spatially symmetric sync state (or its extension, which we also refer to as a ``sync'' state\footnote{When oscillator frequency dispersion $D=0$, a fully synchronized spatially symmetric state $r_1=r_2=1$ exists and is stable.  When $D>0$, perfect synchrony no longer exists, but a spatially symmetric state with $r_1=r_2$ continues to exist. This is the analog of the full sync state, and as is typical in the traditional Kuramoto model, we refer to it as the ``sync'' state even though it does not correspond to perfect synchrony, and even though the order can become quite small if $D$ is large. Any state without spatial symmetry, i.e., with $r_1 \neq r_2$, we refer to as a chimera state.}) to decrease due to the heterogeneity among oscillator natural frequencies. This is apparent from the traditional Kuramoto model, as increasing the dispersion in natural frequencies results in a smaller fraction of oscillators becoming phase-locked. The same phenomenon happens here, as shown in Fig.~\ref{fig:bifurcationasDinc}: moving from the top panel to the bottom, the sync branch lowers as $D$ is increased.  It also ceases to be a horizontal line when $D>0$, and new intersections with the saddle-node branches of solution give rise to two pitchfork bifurcations, one supercritical  and the other subcritical.

As $D$ increases further, the subcritical pitchfork and two saddle-nodes collide leaving behind a second supercritical pitchfork bifurcation.  The third panel in Fig.~\ref{fig:bifurcationasDinc} demonstrates this, and corresponds to a parametric path that intersects the pitchfork balloon without intersecting the saddle-node surface. 

We have now found a connection between the sync state and the stable chimera state via a pitchfork bifurcation! This connection is not evident in the $D=0$ case: the interesting behavior becomes compressed to the $\beta=0$ axis, where the system is integrable \cite{WATANABE1994}. The singular perturbation $D=0 \to D>0$ is necessary to reveal the ``hidden'' pitchfork bifurcations; the same pitchfork concept lies at the heart of many physical systems that spontaneously break symmetry, e.g., buckling in beams, magnetic interactions (Ising model), first-order phase transitions in statistical mechanics, etc.~(see, e.g., \cite{Strogatz1994}).


In Fig.~\ref{fig:chimera_wedge}, the region inside the pitchfork balloon is the region where only the chimera state is stable, and the region between the pitchfork balloon and the saddle-node surface is the region of bistability. The bistable region grows to occupy $100\%$ when $D \to 0$, which explains why chimera states were observed to coexist with a stable sync state in this system with identical oscillators \cite{Abrams2008}.

\textbf{Perturbation theory.}
Motivated by our computational results regarding the geometry of the bifurcation surfaces, we wish to obtain analytical expressions for these surfaces, at least in the limit where parameter values are small.

We start by trying to identify a path through the origin that remains exclusively in one region of the parameter space partition shown in Fig.~\ref{fig:chimera_wedge}; that is, we want to find a parametric path that passes through the origin and doesn't cross either the pitchfork or saddle-node surface.  The parameter bounds on such a path should then correspond to the boundaries we wish to identify.

A straight line path fails, since it exits the chimera wedge near the origin (recall that the wedge ``pinches off'' to the origin in the $D$ direction---see the inset of Fig.~\ref{fig:Aslicechimerawedge}).  That means that, moving along a straight-line path toward the origin, a system initialized in the chimera state would necessarily switch to the sync state before it reached the origin. 

Instead, we find that a path where $D$ scales quadratically works as desired: a system initialized in the chimera state can be continued along such a path arbitrarily close to the origin. We will use this path to seek chimera state solutions to system \eqref{eq:zdyn} in the perturbative limit where $\beta$, $A$, and $D$ are all small.  
%
We thus impose the parameter scaling \{$\beta = k_{\beta} \varepsilon$, $A = k_{A} \varepsilon$, $D = k_{D} \varepsilon^{2}$\} together with the ansatz \{$r_{1} = r_{1,0} + r_{1,1} \varepsilon$, $r_{2} = r_{2,0} + r_{2,1} \varepsilon$, $\phi = \phi_{0} + \phi_{1} \varepsilon$\}, and look for equilibria of system \eqref{eq:zdyn} truncated to successive orders of $\varepsilon$.  

At the lowest order, i.e. $\varepsilon = 0$, we find $r_{1,0} = 1, \ r_{2,0} = 1$. 
Plugging in this zeroth order solution and solving the equations at first order in $\varepsilon$, we find $\phi_{0} = 0$. 
Substituting the zeroth and first order solutions in \eqref{eq:zdyn} and solving them at the second order in $\varepsilon$, we obtain a quartic equation in $\phi_{1}$ and explicit solutions for $r_{1,1}$ and $r_{2,1}$ in terms of its roots:
\begin{subequations}
	\begin{multline} \label{eq:phi1quartic}
		k_{\beta} \phi_{1}^{4} - ( 8 k_{\beta}^{3} - 4 k_{A} k_{D} ) \phi_{1}^{2} + \\
		16 k_{\beta}^{5} - 16 k_{A} k_{\beta}^{2} k_{D} + 16 k_{\beta} k_{D}^{2} = 0.
	\end{multline}
	\vspace{-7mm}
	\begin{equation}
		r_{1,1} = \frac{-k_{D}}{k_{\beta} + \phi_{1}/2},	
	\end{equation}
	\vspace{-7mm}
	\begin{equation}
		r_{2,1} = \frac{-k_{D}}{k_{\beta} - \phi_{1}/2},
	\end{equation}
\end{subequations}

Eq.~\eqref{eq:phi1quartic} is a quadratic in $\phi_{1}^{2}$, and imposing our expectation of real-valued solutions implies that two constraints (which will define our bifurcation surfaces) must be satisfied. The solutions of Eqn.~\eqref{eq:phi1quartic} after simplification are $\phi_{1} = \pm k_{\beta}^{-1/2} [2 k_{\beta} ^{3} \pm k_{d} \sqrt{k_{A}^{2} - 4 k_{\beta}^{2}} - k_{A} k_{D}]^{1/2}$. The first constraint is $k_{A}^{2} - 4 k_{\beta}^{2} > 0$ which gives us the saddle-node surface and is consistent with the expression obtained by Abrams, et al.~\cite{Abrams2008}, i.e. $A - 2\beta = 0$.

\begin{figure}[thb]
    \centering
    \vspace{-2mm}
    \includegraphics[width=\columnwidth]{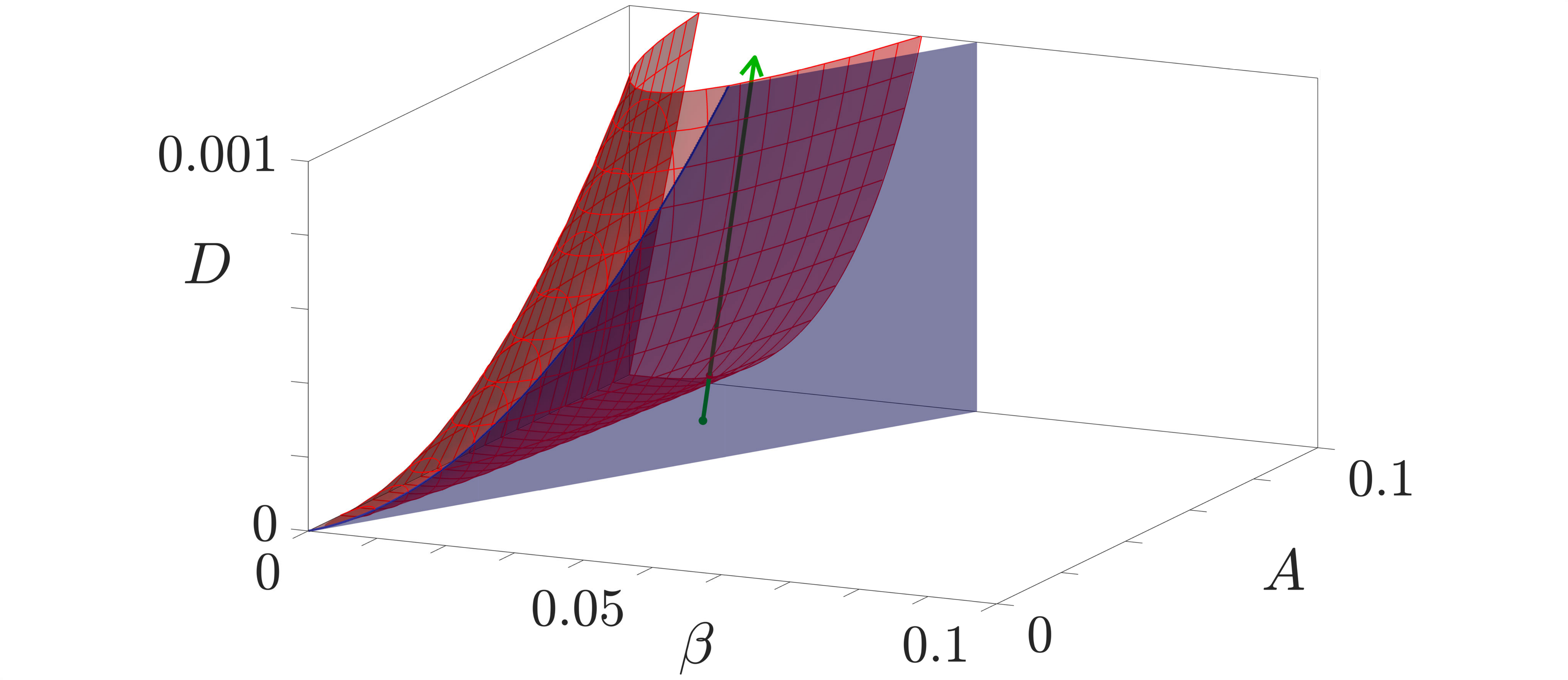}
    \vspace{-7mm}
    \caption{Three-dimensional bifurcation surfaces for equilibria of Eqs.~\eqref{eq:zdyn} in the perturbative limit $\beta, A, D \ll 0$.  Red curved surface: Eq.~\eqref{eq:pitchfork}; blue planar surface: saddle-node surface. Green line: path between chimera state at $(\beta, A, D)=(0.02, 0.08, 0)$ and Kuramoto model state at $(\beta,A,D) = (\pi/2,0,0.2)$ (beyond axis limits).}
    \label{fig:chimera_wedge_theory}
\end{figure}

The second constraint is a pair of inequalities
which gives us the complete pitchfork balloon
\begin{equation}
	2\beta^{3} \pm \sqrt{A^{2}-4\beta^{2}} \ D - A D = 0.
	\label{eq:pitchfork}
\end{equation}
See Fig.~\ref{fig:chimera_wedge_theory} for a plot of these constraint surfaces.
%

We also obtain closed form expressions for stable and unstable spatially symmetric synchronous states. To find these sync states, we plug $r_{1} = r_{2} = r$ into Eqs.~\eqref{eq:zdyn}, set time derivatives to zero, and solve to get the stable sync state (which has $\phi=0$) as
\begin{equation}
	r = \sqrt{1 - \frac{2 D}{\sin \beta}},
\end{equation}
and the unstable sync state (which has $\phi=\pi$) as 
\begin{equation}
	r = \sqrt{1 - \frac{2 D}{A \sin \beta}}.
\end{equation}
These equations imply yet more constraints, namely $\sin \beta > 2 D$ and $A \sin \beta > 2 D$, which specify the regions where stable and unstable sync states exist.

\textbf{A different type of connection.}
Having already obtained a connection between the sync branch and the chimera branch in Fig.~\ref{fig:bifurcationasDinc}, we now explicitly examine the connection between the Kuramoto model and the two-cluster chimera state model. 
%
\begin{figure}[tbh]
    \centering
    \vspace{0mm}
    \includegraphics[width=\columnwidth]{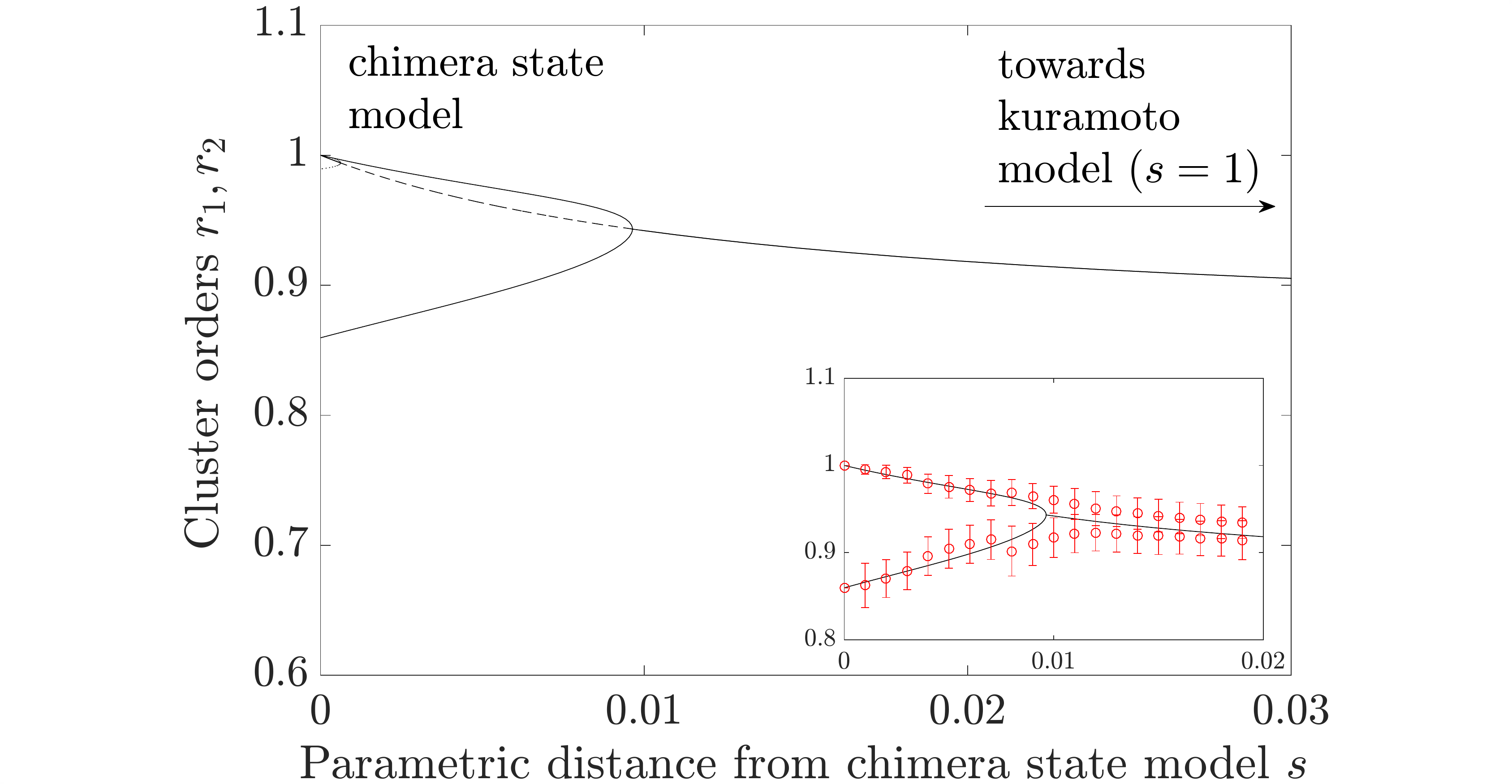}
	\vspace{-6mm}
    \caption{Cluster orders $r_1$ and $r_{2}$ vs distance in parameter space from chimera state $s$ at $(\beta, A, D) = (0.02, 0.08, 0)$ via numerical continuation. Path in parameter space is a straight line connecting the above chimera state to the ``traditional'' Kuramoto model state as $(\beta, A, D) = (\pi/2, 0, 0.2)$. Solid lines: stable branches; dashed or dotted lines: unstable branches. Inset: results from numerical simulation of system \eqref{eq:twoclusterode} with $N_{1} = N_{2} = 256$, with natural frequencies drawn randomly from a Lorentzian distribution $g(\omega)$, with a total integration time of $1500 s$, and overlaid on stable branches from numerical continuation. Circles: order parameters averaged over final $750 s$; error bars: standard deviation of order parameter over final $750 s$. Note that, where multiple branches exist, pairs correspond to interchange symmetry $(r_1=a, r_2=b) \to (r_1=b, r_2=a)$.}
    \label{fig:D2}
\end{figure}
%
%

Motivated by our understanding of the dynamics from the 3-D bifurcation plot, we choose a straight line path from the ``traditional'' Kuramoto model at $(\beta, A, D) = (\pi/2, 0, 0.2)$ (global coupling without phase lag among non-identical oscillators) to a specific chimera state model $(0.02, 0.08, 0)$ (nonzero coupling disparity with phase lag among identical oscillators). This path, shown in Fig.~\ref{fig:chimera_wedge_theory}, intersects only the pitchfork balloon, crossing the surface twice, and thus undergoing two pitchfork bifurcations (note that only one crossing is visible in the figure). 

Starting from the Kuramoto model, as we enter the pitchfork balloon, the stable chimera state branches off and the symmetric sync state loses stability.  This is visible moving along the curve from right to left in Fig.~\ref{fig:D2}, where $s$ indicates the distance in parameter space. Near $s = 0$, we see that there is a small region of bistability corresponding to the tiny region just under the pitchfork balloon. 

The key point of this analysis is to demonstrate a simple, intuitive aspect of the chimera state in this context: a standard pitchfork bifurcation off of the well-understood Kuramoto sync state leads to its appearance.  In this construction, it is not even bistable with the Kuramoto sync state.

\textbf{Conclusions.}
For the ``two-cluster'' system, we have demonstrated that the chimera state emerges from a completely symmetric partially synchronized state familiar from the traditional Kuramoto model. It appears via a pitchfork bifurcation in which symmetry is broken so that oscillators in one cluster synchronize to a greater extent than oscillators in the other.


Future work might build on this insight so that other puzzling aspects of the chimera state can be made clear. In particular, chimera states in variable amplitude oscillators and in systems with inertia remain poorly understood.  Furthermore, we speculate that the connection between ``spot'' and ``spiral'' chimera states in two and higher dimensions might be understandable with an approach like that we use here. For the system on a ring with a nonlocal coupling kernel, we suspect that an approach similar to Laing's \cite{Laing2009} will allow for explicit analysis of the saddle-node and pitchfork surfaces.

\begin{acknowledgments}
The authors thank Carlo Laing for helpful correspondence and Gokul Nair for helpful conversations.
\end{acknowledgments}







\bibliographystyle{apsrev4-1}

%

\end{document}